\documentclass[USenglish, twocolumn]{article}	

\usepackage[utf8]{inputenc}				
\usepackage[big,online]{dgruyter}	
\usepackage{lmodern} 
\usepackage{microtype}
\usepackage[numbers,square,sort&compress]{natbib}

\theoremstyle{dgthm}

\theoremstyle{dgdef}

\begin{document}

	\articletype{Research Article}
  \startpage{1}
  \aop

\title{Polaritonic Tamm states induced by cavity photons}
\runningtitle{Tamm Polaritons}


\author[1]{C. A. Downing}
\author*[2]{L. Mart\'{i}n-Moreno}


\runningauthor{Downing and Mart\'{i}n-Moreno}
\affil[1]{\protect\raggedright 
Instituto de Nanociencia y Materiales de Arag\'{o}n (INMA), CSIC-Universidad de Zaragoza, Zaragoza 50009, Spain and Departamento de F\'{i}sica de la Materia Condensada, Universidad de Zaragoza, Zaragoza 50009, Spain}
\affil[2]{\protect\raggedright 
Instituto de Nanociencia y Materiales de Arag\'{o}n (INMA), CSIC-Universidad de Zaragoza, Zaragoza 50009, Spain and Departamento de F\'{i}sica de la Materia Condensada, Universidad de Zaragoza, Zaragoza 50009, Spain, e-mail: lmm@unizar.es}


	
\abstract{We consider a periodic chain of oscillating dipoles, interacting via long-range dipole-dipole interactions, embedded inside a cuboid cavity waveguide. We show that the mixing between the dipolar excitations and cavity photons into polaritons can lead to the appearance of new states localized at the ends of the dipolar chain, which are reminiscent of Tamm surface states found in electronic systems. A crucial requirement for the formation of polaritonic Tamm states is that the cavity cross-section is above a critical size. Above this threshold, the degree of localization of the Tamm states is highly dependent on the cavity size, since their participation ratio scales linearly with the cavity cross-sectional area. Our findings may be important for quantum confinement effects in one-dimensional systems with strong light-matter coupling.}




\maketitle

	
\section{Introduction}
\label{intro}

In 1932, Tamm showed the existence of surface states in a one-dimensional (1-D) crystal lattice~\cite{Tamm1932}, due to the abrupt termination of the periodic crystal at an interfacing surface, such as the vacuum~\cite{Fowler1933, Shockley1939, Forstmann1993}. This result highlighted a surprising failure of the theory of a periodic potential with cyclic boundary conditions at the elementary level of the electronic bandstructure, despite its great utility in explaining the bulk properties of solids~\cite{Davison1996, Ren2005}. Tamm surface states have since been shown to have profound consequences for the rich field of surface science, including for photoluminescence in mesoscopic systems~\cite{Ohno1990}, photocurrents in superlattices~\cite{Ohno1990b}, and the absorption spectra of molecular chains~\cite{Agranovich2000, Schmidt2002, Hoffmann2003}.

Over the last two decades, various theories of Tamm states in the latest condensed matter systems have been developed \cite{Vinogradov2010}. For example, with exciton-polaritons in multilayer dielectric structures~\cite{Kavokin2005}, with plasmons at at the boundary between a metal and a dielectric Bragg mirror~\cite{Kaliteevski2007}, and with phonons in graphene nanoribbons~\cite{Savin2010}. Pioneering experimental work has seen the observance of Tamm states in magnetophotonic structures~\cite{Goto2008}, in organic dye-doped polymer layers~\cite{Nunez2016}, and latterly in photonic crystals~\cite{Fecteau2018, Fecteau2019, Nakata2020}.

Due to the rise of topological physics in photonics and plasmonics~\cite{Slobo2015, Ling2015, Downing2017, Pocock2018, Downing2018, Gutierrez2018}, there is an ongoing interest in finding and classifying unconventional light-matter states. Indeed, the latest advances in topological matter have been made in photon-based systems, leading to the rapidly expanding subfield of topological nanophotonics~\cite{Lu2014, Khanikaev2017, Sun2017, Martinez2018, Ozawa2019, Rider2019, Ota2019}. It is therefore crucial to also classify and understand surface states of a non-topological origin, such as Tamm-like edge states, in systems with strong light-matter coupling. Indeed, there are recent experimental studies of polariton micropillars where the localization of both topologically trivial and topologically non-trivial modes are examined in detail \cite{Jean2017, Whittaker2019}.

\begin{figure}[tb]
 \includegraphics[width=1.0\linewidth]{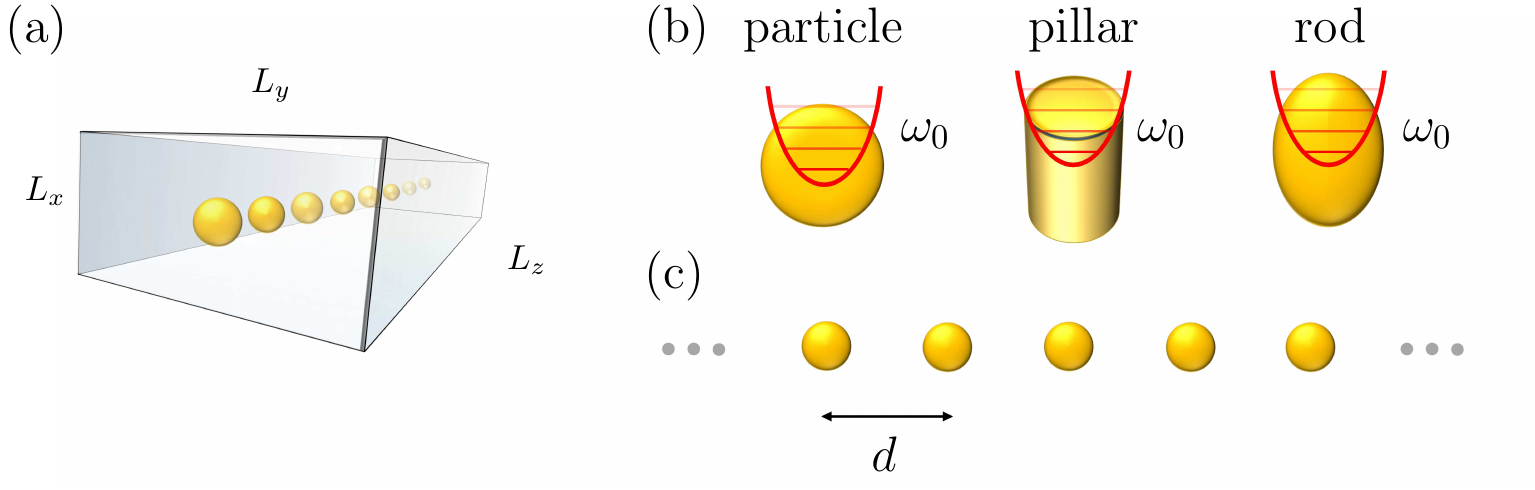}
 \caption{ Panel (a): a sketch of our system: a chain of dipoles embedded inside a cuboid cavity waveguide, of dimensions $L_x \times L_y \times L_z$. Panel (b): Each dipole is modeled as a harmonic oscillator of resonance frequency $\omega_0$. It can be realized by the Mie resonance in a dielectric nanoparticle, spin waves in a magnetic micropillar or localized surface plasmons in a metallic nanorod. Panel (c): the long chain of $\mathcal{N} \gg 1$ oscillating dipoles, regularly spaced by the center-to-center separation $d$.  }
 \label{sketch}
\end{figure}

In this work, we consider a nanophotonic system which exhibits Tamm-like edge states: a 1-D chain of regularly spaced nanoresonators, coupled via dipole-dipole interactions, which are housed inside a cavity waveguide [see Fig.~\ref{sketch}~(a)]. The linear dipolar chain is of some importance, since it is a simple system where one may study the subwavelength transportation of energy and information~\cite{Brongersma2000, Maier2003, Park2004, Brandstetter2016, Downing2018b}. In our theory, we place the dipolar chain inside a cuboid cavity in order to study the effect of controllable light-matter interactions. Modulating the cross-sectional area of the cavity allows one to tune both the light-matter coupling strength and the light-matter detuning. In the strong coupling regime the dipolar excitations in the resonator chain hybridize with the cavity photons to form polaritonic excitations~\cite{Ameling2013, Ginzburg2016}. The resulting polaritons, which display half-light and half-matter properties, can lead to the emergence of a highly localized edge state of a non-topological origin: a Tamm-like state. Notably, neither the dipolar chain nor the cavity photons display Tamm states when the light-matter interaction is switched off. We discuss the properties of the emergent polaritonic Tamm state, including how its creation requires the cavity cross-section to be above a critical size, and how its localization properties scale with the cavity cross-sectional area.

The presented theory of a chain of oscillating dipoles embedded inside a cuboid cavity may be realized in a wide range of dipolar systems, as alluded to in the sketches in Fig.~\ref{sketch}~(b). At the subwavelength scale, exploiting the Mie resonances in a chain of dielectric nanoparticles is a promising option, since the system does not suffer from high losses, and is hence ideal for energy transportation~\cite{Savelev2016, Bakker2017, Koshelev2020}. Localized surface plasmons hosted by metallic nanoparticles are another accessible platform~\cite{Barrow2014, Gur2018, Rekola2018}, and there are several recent experimental studies of plasmonic nanoparticles in cavity geometries~\cite{Barth2010, Huang2011, Schmidt2012, Schmidt2018}. Exciting spin waves in magnetic microspheres is another appealing possibility~\cite{Pirmoradian2018}, since cavity magnons have been well studied experimentally in recent years~\cite{Zhang2014}. Finally, implementations with Rydberg~\cite{Browaeys2016, Leseleuc2019} and ultracold atoms~\cite{Weimer2012, Cooper2019}, as well as helical resonators~\cite{Mann2018}, are also viable settings for the versatile theory presented here.

The rest of the manuscript is organized as follows: we describe our model in Sec.~\ref{model}, we unveil the polaritonic Tamm states in Sec.~\ref{sec:topo}, and we draw some conclusions in Sec.~\ref{conc}. The supplemental material contains additional theoretical details.


\section{Model}
\label{model}

The Hamiltonian of a chain of oscillating dipoles embedded inside a cavity reads~\cite{Craig1984, Salam2010, Downing2019}
\begin{equation}
\label{eq:Ham}
H = H_{\mathrm{dp}} + H_\mathrm{ph} + H_{\mathrm{dp}\textrm{-}\mathrm{ph}},
\end{equation}
accounting for the dipolar dynamics, the cavity photons and the light-matter coupling respectively. Importantly, the couplings in $H_{\mathrm{dp}}$ go beyond the nearest-neighbor approximation~\cite{Downing2018b, Downing2018}, which is essential for a proper treatment of the type of Tamm states discussed in this system.

We sketch in Fig.~\ref{sketch}~(c) the model of our system: a 1-D array of dipoles, regularly spaced at the interval $d$, which is encased inside a cuboid cavity of dimensions $L_x \times L_y \times L_z$ [see panel~(a)]. Tuning the size of the cavity cross-sectional area ($L_x \times L_y$) modulates both the light-matter coupling strength and the light-matter detuning, such that polariton excitations may be formed by the mixing between the cavity photons and dipolar excitations [which are generally treated as harmonic oscillators, see Fig.~\ref{sketch}~(b) for some physical realizations].


\subsection{\label{jambon123}Dipolar Hamiltonian}

\begin{figure*}[tb]
 \includegraphics[width=1.0\linewidth]{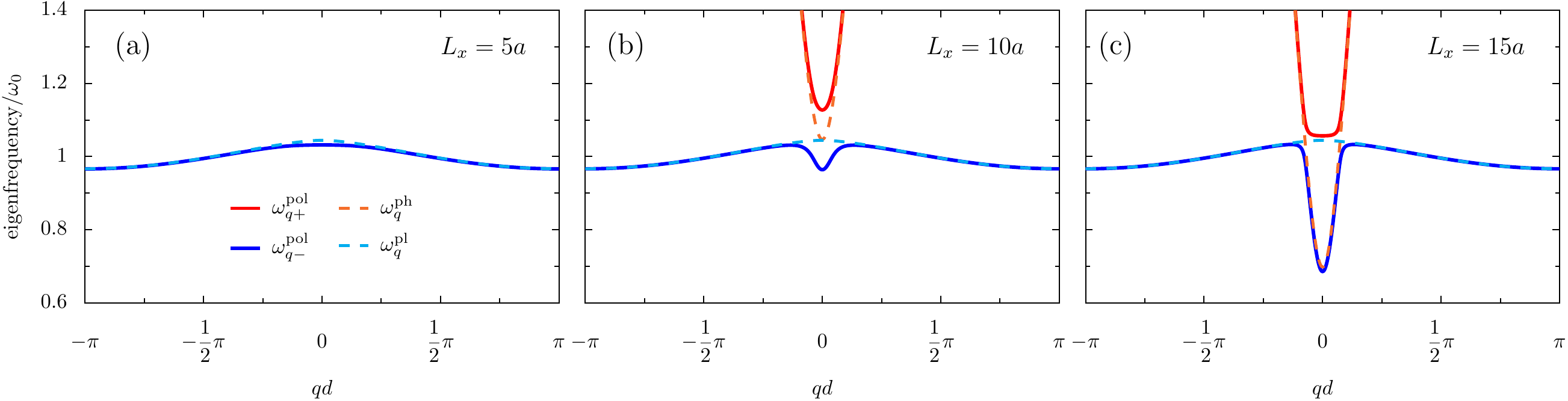}
 \caption{ The polariton dispersion $\omega_{q \tau}^{\mathrm{pol}}$ in the first Brillouin zone [cf. Eq.~\eqref{eq:Ham_polly_spec}], for the cavity heights (a) $L_x = 5 a$, (b) $L_x = 10 a$, and (c) $L_x = 15 a$. The upper (lower) polaritons with $\tau = + (-)$ are denoted by solid blue (red) lines. The uncoupled photonic (dipolar) dispersions $\omega_{q}^{\mathrm{ph}}$ ($\omega_{q}^{\mathrm{dp}}$) are shown as dashed cyan (orange) lines [cf. Eq.~\eqref{eq:LM_disp} and Eq.~\eqref{eq:plspectrum}]. In the figure, the inter-dipole separation $d = 3a$, the dipole strength $\omega_0 c/a = 1/10$, and the cavity aspect ratio $L_y = 3 L_x$. }
 \label{bandstructure}
\end{figure*}

The dipolar Hamiltonian [$H_{\mathrm{dp}}$ in Eq.~\eqref{eq:Ham}] describes a linear chain of $\mathcal{N} \gg 1$ dipoles [cf. Fig.~\ref{sketch}~(c)], oscillating in the $\hat{x}$-direction with transverse polarization ($\uparrow \uparrow \uparrow \cdots$), and coupled to each other via dipole-dipole interactions. Setting $\hbar = 1$ throughout, this Hamiltonian reads (see Refs.~\cite{Brandstetter2016, Downing2018b, SuppMatt} for details)
\begin{equation}
\label{eq:Ham_transformed}
  H_{\mathrm{dp}} =  \sum_q \bigg\{ \omega_0 \: b_{q}^{\dagger} b_{q} 
+ \frac{\Omega}{2} f_q  \left[ b_{q}^{\dagger} \left( b_q + b_{-q}^{\dagger} \right) + \mathrm{h. c.} \right] \bigg\}, 
 \end{equation}
where the bosonic creation (annihilation) operator $b_{q}^{\dagger}$ ($b_q$) creates (destroys) a dipolar excitation of wavevector $q$, where $q \in [-\pi/d, +\pi/d]$ spans the first Brillouin zone. The dipolar resonance frequency of a single dipole is $\omega_0$, which is associated with the length scale $a$~\cite{SuppMatt}. The weak dipolar coupling constant $\Omega \ll \omega_0$ reads $\Omega =  (\omega_0/2) (a/d)^3$, exhibiting the inverse-cubic dependence characteristic of dipole-dipole interactions, and $d$ is the center-to-center separation between the dipoles~\cite{SuppMatt}. In Eq.~\eqref{eq:Ham_transformed}, we have introduced the lattice sum $f_{q} = 2 \sum_{n=1}^\infty \cos{(nqd)}/n^3 = 2 \mathrm{Cl}_3 (qd)$, where $\mathrm{Cl}_s (z)$ is the Clausen function of order $s$. Crucially, $f_{q}$ takes into account long-range interactions between all of the resonators, which is known to be important in dipolar systems~\cite{Downing2018b}. Significantly, going beyond the nearest-neighbor approximation also changes the constraints at the edge of the chain from standard hard-wall boundary conditions. Namely, the edge resonators $1$ and $\mathcal{N}$ do not just feel the penultimate resonators $2$ and $\mathcal{N}-1$ but also those in the bulk. When considering the nearest-neighbor ($\mathrm{nn}$) coupling approximation, one should replace the lattice sum $f_{q}$ with the standard result $f_{q}^{\mathrm{nn}} = 2 \cos (qd)$.

After ignoring the counter-rotating terms in Eq.~\eqref{eq:Ham_transformed} [see Ref.~\cite{SuppMatt} for the full treatment], the eigenfrequencies $\omega_{q}^{\mathrm{dp}}$ of the collective dipolar modes follow immediately as 
\begin{equation}
\label{eq:plspectrum}
\omega_{q}^{\mathrm{dp}} = \omega_0 + \Omega f_q.
\end{equation}
Equation~\eqref{eq:plspectrum} describes the usual space quantization of eigenfrequencies into a solitary band. Within the nearest-neighbor coupling approximation $f_{q} \to f_{q}^{\mathrm{nn}}$, the spectrum of Eq.~\eqref{eq:plspectrum} reduces to the more familiar cosine expression, $\omega_{q}^{\mathrm{dp, nn}} = \omega_0 + 2 \Omega \cos (qd)$. The most noticeable impact of the above approximation, at the level of the continuum bandstructure, is a reduction in the dipolar bandwidth $B^{\mathrm{dp}}$ to the nearest-neighbor value $B^{\mathrm{dp, nn}} = 4 \Omega$. The higher ``all-coupling'' value, which follows from Eq.~\eqref{eq:plspectrum}, is $B^{\mathrm{dp}} = (7/2) \zeta (3) \Omega \simeq 4.21 \Omega$. Here $\zeta (3) = 1.202 \ldots$ is Ap\'{e}ry's constant and $\zeta (s)$ is the Riemann zeta function.


\subsection{\label{jambon456}Polaritonic Hamiltonian}

The photonic Hamiltonian [$H_{\mathrm{ph}}$ in Eq.~\eqref{eq:Ham}] describes the cavity photons inside the long cuboid cavity of dimensions $L_z \gg  L_y > L_x$ [see Fig.~\ref{sketch}~(a)]. In terms of the photonic creation (annihilation) operator $c_{q}^{\dagger}$ ($c_q$), it reads (see Refs.~\cite{Milonni1994, Kakazu1994, SuppMatt} for details)
\begin{equation}
\label{eq:H_ph_cavity}
 H_{\mathrm{ph}} = \sum_{q} \omega_q^{\mathrm{ph}} c_{q}^{\dagger} c_{q}.
\end{equation}
where the cavity photon dispersion $\omega_q^{\mathrm{ph}}$ is given by
\begin{equation}
\label{eq:LM_disp}
\omega_q^{\mathrm{ph}} = c \sqrt{ q^2 + \left( \frac{\pi}{L_y} \right)^2  },
\end{equation}
where $c$ is the speed of light in vacuum. The cavity width is $L_y$ and the cavity aspect ratio $L_y > L_x$, such that only the photonic band of Eq.~\eqref{eq:LM_disp} is relevant for the problem. The full light-matter coupling Hamiltonian [$H_{\mathrm{dp}\textrm{-}\mathrm{ph}}$ in Eq.~\eqref{eq:Ham}] reads [see Ref.~\cite{SuppMatt} for the derivation]  
\begin{align}
\label{eq:light_matter_coupling}
H_{\mathrm{dp}\textrm{-}\mathrm{ph}} =& \sum_{ q} \bigg\{ \mathrm{i} \xi_{ q} \Big[   b_{q}^{\dagger}  \left( c_{q} + c_{ -q}^{\dagger} \right) - \mathrm{h. c.}  \Big] \nonumber \\
&\quad + \frac{\xi_{ q}^2}{\omega_0} \Big[ c_{ q}^{\dagger} \left( c_{ q} + c_{ -q}^{\dagger} \right) + \mathrm{h. c.} \Big] \bigg\},
\end{align}
where we have introduced the light-matter coupling constant 
\begin{equation}
\label{eq:LM_Coupling}
\xi_{q} = \omega_0 \left(  \frac{2 \pi a^3}{L_x L_y d} \frac{\omega_0}{\omega_q^{\mathrm{ph}}} \right)^{1/2}.
\end{equation}
The diamagnetic term [on the second line of Eq.~\eqref{eq:light_matter_coupling}] simply leads to a renormalization of the photon dispersion $\omega_{q}^{\mathrm{ph}}$, as defined in Eq.~\eqref{eq:LM_disp}, into
\begin{equation}
\label{eq:photyb}
\tilde{\omega}_{q}^{\mathrm{ph}} = \omega_{q}^{\mathrm{ph}} + \frac{2\xi_{q}^2}{\omega_0},
\end{equation}
a shift which can be safely disregarded throughout this work, since it only leads to small quantitative changes to the results presented here. The paramagnetic term [on the first line of Eq.~\eqref{eq:light_matter_coupling}] is important and gives rise to the formation of polaritonic excitations.

Ignoring counter-rotating terms in the polaritonic Hamiltonian [formed by Eq.~\eqref{eq:Ham_transformed}, Eq.~\eqref{eq:H_ph_cavity} and Eq.~\eqref{eq:light_matter_coupling}], we may write the resulting rotating wave approximation (RWA) polaritonic Hamiltonian as follows
\begin{equation}
\label{eq:Ham_polly_other}
H_{\mathrm{pol}}^{\mathrm{RWA}} = \sum_{q} \hat{\psi^{\dagger}} \mathcal{H}_{\mathrm{pol}}^{\mathrm{RWA}} \hat{\psi},
\quad
\mathcal{H}_{\mathrm{pol}}^{\mathrm{RWA}} =
\begin{pmatrix}
\omega_{q}^{\mathrm{dp}} && \mathrm{i} \xi_q \\
-\mathrm{i} \xi_q && \omega_{q}^{\mathrm{ph}}
\end{pmatrix},
\end{equation}
where the polaritonic Bloch Hamiltonian is $\mathcal{H}_{\mathrm{pol}}^{\mathrm{RWA}}$, and where we used the basis $\hat{\psi} = ( b_q, c_q )$. We arrive by bosonic Bogoliubov transformation at the diagonal form of Eq.~\eqref{eq:Ham_polly_other}
\begin{equation}
\label{eq:Ham_polly}
H_{\mathrm{pol}}^{\mathrm{RWA}} = \sum_{q \tau} \omega_{q \tau}^{\mathrm{pol}} \beta_{q \tau}^{\dagger} \beta_{q \tau},
\end{equation}
where the index $\tau = \pm$ labels the upper and lower polariton bands. The polariton dispersion $\omega_{q \tau}^{\mathrm{pol}}$ in Eq.~\eqref{eq:Ham_polly} reads
\begin{equation}
\label{eq:Ham_polly_spec}
\omega_{q \tau}^{\mathrm{pol}} = \bar{\omega}_q + \tau \Omega_q, 
\end{equation}
where the average frequency of the uncoupled dispersions $\bar{\omega}_q$, the effective coupling constant $\Omega_q$, and the light-matter detuning $\Delta_q$ are given by
\begin{align}
\label{eq:Ham_polly_spec_2}
\bar{\omega}_q = \tfrac{1}{2} &\left( \omega_{q}^{\mathrm{ph}} + \omega_{q}^{\mathrm{dp}}  \right),
\quad \Omega_q = \sqrt{ \xi_q^2 + \Delta_q^2 }, \nonumber \\
&\Delta_q = \tfrac{1}{2} \left( \omega_{q}^{\mathrm{ph}} - \omega_{q}^{\mathrm{dp}} \right).
\end{align}
The bosonic Bogoliubov operators $\beta_{q \tau}$ in Eq.~\eqref{eq:Ham_polly} are defined by
\begin{equation}
\label{eq:Bog2433}
 \beta_{q +} =  \sin \theta_q  b_{q} - \mathrm{i} \cos \theta_q  c_{q}, 
 \beta_{q -} =  \cos  \theta_q  b_{q} + \mathrm{i} \sin  \theta_q  c_{q},
 \end{equation}
where the Bogoliubov coefficients are
\begin{equation}
\label{eq:Bog_coe434ff23}
 \cos \theta_q  = \frac{1}{\sqrt{2}} \left( 1 + \frac{\Delta_q}{\Omega_q} \right)^{\tfrac{1}{2}}, 
 \sin \theta_q  = \frac{1}{\sqrt{2}} \left( 1 - \frac{\Delta_q}{\Omega_q} \right)^{\tfrac{1}{2}},
\end{equation}
in terms of the quantities defined in Eq.~\eqref{eq:Ham_polly_spec_2}.

We plot in Fig.~\ref{bandstructure} the polariton dispersion of Eq.~\eqref{eq:Ham_polly_spec} for the cavity heights $L_x = \{  5 a, 10a, 15a \}$ in panels (a), (b) and (c) respectively, where the cavity aspect ratio is fixed at $L_y = 3 L_x$, and the inter-dipole separation at $d= 3a$. With increasing cavity cross-sectional area in going from panel (a) to (b) to (c), the light-matter detuning $\Delta_q$ is reduced [cf. Eq.~\eqref{eq:Ham_polly_spec_2}], leading to increasingly noticeable deviations of the polariton bands (solid lines) from the uncoupled dispersions (dashed lines). The upper (lower) polariton band is given by the red (blue) lines. The photonic bandstructure is denoted by orange lines, while the dipolar bands are in cyan. Notably, in panel (a) only a single polaritonic band is visible on this scale, since the (mostly photonic) upper polariton band lies significantly above the frequency scale of $\omega_0$. Panel (b) demonstrates the strong coupling regime and its associated highly reconstructed polariton dispersion, while panel (c) displays the usual band anti-crossing behavior as the detuning is further reduced.

The Bogoliubov operators of Eq.~\eqref{eq:Bog2433} imply the pair of polaritonic Bloch spinors $\psi_{q +} = ( \sin \theta_q,  - \mathrm{i} \cos \theta_q )^{\mathrm{T}}$ and $\psi_{q -} = ( \cos  \theta_q, \mathrm{i} \sin  \theta_q )^{\mathrm{T}}$. Notably, unlike the celebrated spinors describing excitations in some topologically nontrivial systems~\cite{Slobo2015, Ling2015, Downing2017, Pocock2018, Downing2018, Gutierrez2018, Henriques2020}, there is not a $q$-dependent phase factor difference (like $\mathrm{e}^{\mathrm{i} \delta_q}$) between the upper and lower components of each individual spinor $\psi_{q \tau}$. This suggests the absence of any topological physics, which can be confirmed by analyzing the Bloch Hamiltonian. The Hamiltonian of $\mathcal{H}_{\mathrm{pol}}$ in Eq.~\eqref{eq:Ham_polly_other} can be decomposed into a 1-D Dirac-like Hamiltonian
\begin{equation}
\label{eq:Diraccy}
 \mathcal{H}^{\mathrm{pol}}_q = \bar{\omega}_q I_2 - \Delta_q \sigma_z - \xi_q \sigma_y,
\end{equation}
where $\{ \sigma_x, \sigma_y, \sigma_z \}$ are the Pauli matrices, and $I_2$ is the two-dimensional identity matrix. Despite this Dirac mapping, the associated spinors $ \psi_{q \tau}$ indeed lead to a trivial Zak phase of zero~\cite{Berry1984, Zak1989}. This triviality follows from the symmetries of the Bloch Hamiltonian of Eq.~\eqref{eq:Diraccy}, which displays broken inversion ($\sigma_x \mathcal{H}^{\mathrm{pol}}_{-q} \sigma_x \ne \mathcal{H}^{\mathrm{pol}}_{q}$) and chiral ($\sigma_z \mathcal{H}^{\mathrm{pol}}_{q} \sigma_z \ne - \mathcal{H}^{\mathrm{pol}}_{q}$) symmetries~\cite{Asboth2016}. This Zak phase analysis classifies the system as topologically trivial, which hence implies an absence of topologically-protected edge states. This fact motivates us to understand the highly localized, and yet non-topological, states which can nevertheless be supported by this system (as is shown in the next section).

Perhaps surprisingly, the mixing between the dipolar and photonic modes into polaritons also gives rise to the formation of Tamm-like edge states. These localized states are missing in Fig.~\ref{bandstructure}, since their emergence requires a finite system (which precludes the use of periodic boundary conditions). 


\section{Polaritonic Tamm states}
\label{sec:topo}

\begin{figure*}[tb]
 \includegraphics[width=1.0\linewidth]{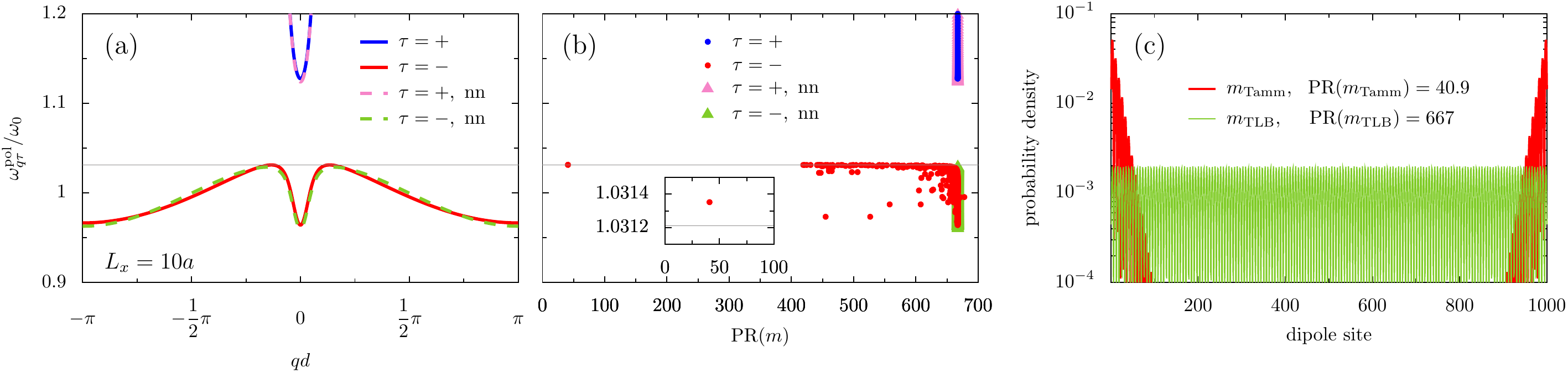}
 \caption{ Panel (a): the polariton dispersion in the first Brillouin zone, with all-neighbor coupling $\omega_{q \tau}^{\mathrm{pol}}$ (nearest-neighbor coupling $\omega_{q \tau}^{\mathrm{pol, nn}}$) [cf. Eq.~\eqref{eq:Ham_polly_spec}]. The upper $\tau = +$ polariton band is denoted by solid blue (dashed pink) lines and the lower $\tau = -$ polariton band is denoted by solid red (dashed green) lines for all (nearest)-neighbor coupling. Horizontal gray line: guide for the eye at the eigenfrequency which corresponds to the top of the bulk band. Panel (b): the polariton eigenfrequencies with all-neighbor coupling $\omega_{m}^{\mathrm{pol}}$ (nearest-neighbor coupling $\omega_{m}^{\mathrm{pol, nn}}$), calculated in real space for a chain of $\mathcal{N} = 1000$ dipoles, as a function of the participation ratio $\mathrm{PR}(m)$, where $m$ labels the eigenstate [cf. Eq.~\eqref{eq:p_r412}]. The color scheme is the same as in panel (a). Inset: a zoom in of the Tamm state, which lies just above the bulk band. Panel (c): the probability density across the dipolar chain, for the polariton eigenstate at the top of the lower polariton band $m_{\mathrm{TLB}}$ (for the Tamm state $m_{\mathrm{Tamm}}$), where nearest (all)-neighbor coupling is given by the thin green (thick red) solid line. In the figure, the inter-dipole separation $d = 3a$, the dipole strength $\omega_0 c/a = 1/10$, the cavity height $L_x = 10 a$, and the cavity aspect ratio $L_y = 3 L_x$. }
 \label{ipr}
\end{figure*}

In order to search for the edge states in our system of a chain of resonators inside a cavity, it is necessary to solve the eigenproblem of Eq.~\eqref{eq:Ham} in real space, thus removing the periodic boundary condition assumption in the Fourier space calculation of the previous section. This procedure leads to the eigenfrequencies $\omega_{m}^{\mathrm{pol}}$ (and $\omega_{m}^{\mathrm{pol, nn}}$ in the nearest-neighbor approximation), where each eigenstate is labelled with the index $m$. Each eigenstate $\psi(m) = (\psi_1, \cdots, \psi_{\mathcal{N}} )$ spans every site in the chain of $\mathcal{N}$ dipoles. The localization of the states may be classified by the participation ratio $\mathrm{PR}(m)$, as defined by~\cite{Bell1970, Thouless1974}
\begin{equation}
\label{eq:p_r412}
 \mathrm{PR}(m) = \frac{ \left( \sum_{n=1}^{\mathcal{N}} | \psi_n(m) |^2 \right)^2 } {\sum_{n=1}^{\mathcal{N}} | \psi_n(m) |^4},
\end{equation}
where the summations are over all of the dipole sites $n$. Extended states residing in the bulk part of the spectrum are characterized by a participation ratio scaling linearly with the system size, and in the nearest-neighbor approximation $\mathrm{PR}(m) \simeq (2/3) \mathcal{N}$~\cite{SuppMatt}. Notably, the participation ratio of edge states does not scale with the system size $\mathcal{N}$. 

In Fig.~\ref{ipr}~(a), we plot the polariton bandstructure from Eq.~\eqref{eq:Ham_polly_spec} with $L_x = 10a$ [cf. Fig.~\ref{bandstructure}~(b)], where all-neighbor coupling $\omega_{q \tau}^{\mathrm{pol}}$ (nearest-neighbor coupling $\omega_{q \tau}^{\mathrm{pol, nn}}$) is denoted by solid lines (dashed lines). The upper (lower) polariton band is red (blue) for all-neighbor coupling, and green (pink) for the nearest-neighbor coupling approximation. The horizontal gray line is a guide for the eye at the eigenfrequency corresponding to the top of the lower polariton band ($\tau = -1$). Clearly, the impact on the continuum bandstructure of going beyond the nearest-neighbor approximation is negligible, perhaps making the appearance of edge states in the corresponding finite system even more surprising.

Using Eq.~\eqref{eq:p_r412}, Fig.~\ref{ipr}~(b) displays the participation ratio $\mathrm{PR}(m)$ for the equivalent problem in real space for a chain of $\mathcal{N} = 1000$ dipoles, and the color scheme is the same as in panel~(a). Strikingly, the participation ratio of the polariton states in the nearest-neighbor coupling case (green and pink triangles) is essentially uniform [$\mathrm{PR}^{\mathrm{nn}}(m) \simeq (2/3) 1000 \simeq 667$], while for the all-neighbor case the participation ratio of the lower polariton band (blue circles) is markedly different, especially near to the top of the lower polariton band ($\mathrm{TLB}$). In particular, the state at the very top of the lower polariton band in the nearest neighbor approximation is associated with $\mathrm{PR}^{\mathrm{nn}} (m_{\mathrm{TLB}}) \simeq 667$, while in the all-coupling case the Tamm state just above this band (which we ascribe with the state index $m_{\mathrm{Tamm}}$) has $\mathrm{PR} (m_{\mathrm{Tamm}}) \simeq 41$ [see the insert in panel (b) for a zoom in on the Tamm state]. This last result suggests a highly localized state, induced by the different boundary conditions in the all-coupling case, as compared to the standard hard-wall conditions in the nearest-neighbor approximation.

\begin{figure*}[tb]
 \includegraphics[width=1.0\linewidth]{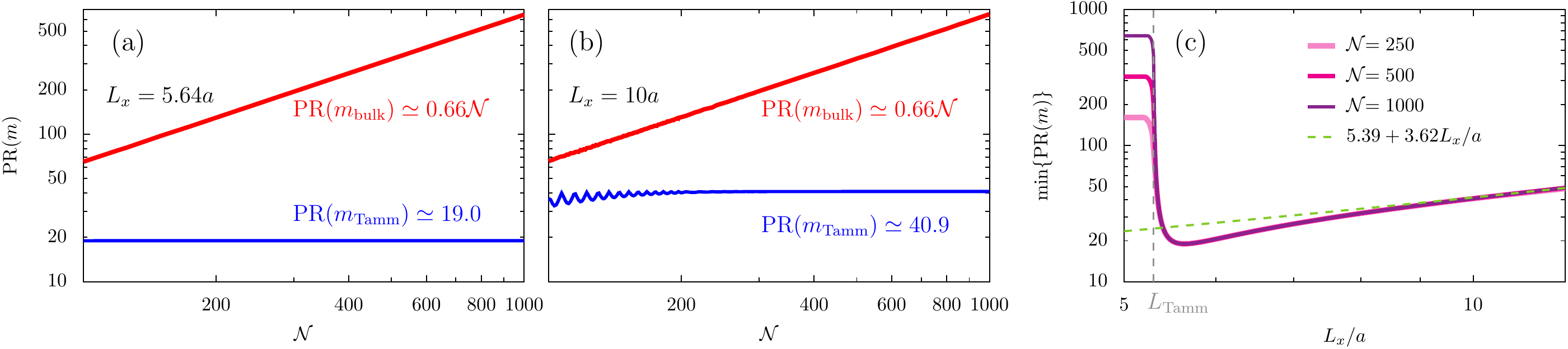}
 \caption{ Panels (a) and (b): the participation ratio $\mathrm{PR}(m)$ as a function of the number of dipoles $\mathcal{N}$ in the chain, where bulk (Tamm) states are denoted by the thick red (thin blue) lines [cf. Eq.~\eqref{eq:p_r412}]. We show results for the cavity heights $L_x = 5.64 a$ [panel (a)] and $L_x = 10 a$ [panel~(b)]. Panel (c): the minimum of the participation ratio $\mathrm{min} \{ \mathrm{PR}(m) \}$ as a function of the reduced cavity height $L_x/a$, calculated for $\mathcal{N} = \{ 250, 500, 1000\}$ dipoles. The linear fitting valid for $L_x \gtrsim 8 a$ is given by the dashed green line. The critical cavity size $L_{\mathrm{Tamm}}$ is denoted by the vertical dashed gray line. In the figure, the inter-dipole separation $d = 3a$, the dipole strength $\omega_0 c/a = 1/10$, and the cavity aspect ratio $L_y = 3 L_x$.}
 \label{finiteeffect}
\end{figure*}

We plot in Fig.~\ref{ipr}~(c) the probability density $|\psi_n|^2$ along the dipolar chain, where the sites are labelled by $n$, for the polariton eigenstate at the top of the lower polariton band, where the nearest (all)-neighbor coupling is given by the thin green (thick red) line and is associated with the index $m_{\mathrm{TLB}}$ ($m_{\mathrm{Tamm}}$). This panel clearly displays the emergence of the Tamm-like edge state in the all-coupling case, induced by (i) the strong light-matter coupling in the cavity, and (ii) the all-coupling boundary conditions. This state is not associated with a topological invariant [see the discussion after Eq.~\eqref{eq:Diraccy}], and so we term it a polaritonic Tamm state. This is in direct analogy with the non-topological surface states studied in solid state physics, which also typically arise in 1-D tight-binding models.

In Fig.~\ref{finiteeffect}~(a)~and~(b), we show the dependence of the participation ratio $\mathrm{PR}(m)$ on the number of dipoles in the chain $\mathcal{N}$, for the cavity heights $L_x = 5.64 a$ in panel~(a) [the reason for this choice will become apparent in what follows] and $L_x = 10 a$ in panel~(b). The results for the Tamm states are denoted by thin blue lines, while the results for the bulk states are represented by thick red lines, and the equation of the line is labelled nearby. These results confirm that the bulk states behave according to the standard formula $\mathrm{PR}(m_{\mathrm{bulk}}) \simeq (2/3) \mathcal{N}$ (see Ref.~\cite{SuppMatt}), and reveals that the exotic state revealed in Fig.~\ref{ipr}~(b) is indeed a highly localized edge state, since it persists with $\mathrm{PR}(m_{\mathrm{Tamm}}) \simeq \mathrm{constant}$ with increasing $\mathcal{N}$. Clearly, the increased cavity height $L_x$ in going from panel (a) to (b) in Fig.~\ref{finiteeffect} has led to an increased participation ratio of the Tamm states, suggesting weaker localization [explicitly a rise from $\mathrm{PR}(m_{\mathrm{Tamm}}) \simeq 19.0$ in panel (a) to $\mathrm{PR}(m_{\mathrm{Tamm}}) \simeq 40.9$ in panel (b)]. This reveals the simple modulation of the cavity cross-sectional area as a tool to control the degree of localization of the edge state (supplementary plots for other cavity sizes are given in Ref.~\cite{SuppMatt}).

We investigate the cavity size-Tamm state relationship in Fig.~\ref{finiteeffect}~(c), which shows the minimum of the participation ratio $\mathrm{min} \{ \mathrm{PR}(m) \}$ as a function of the cavity height $L_x$, for chains of $\mathcal{N} = \{ 250, 500, 1000\}$ dipoles. It exposes the identity of the critical length scale $L_{\mathrm{Tamm}} \simeq 5.30 a$, the cavity height above which the Tamm states first appear in the system. For subcritical cavities ($L_x < L_{\mathrm{Tamm}}$) no edge states are present, as in a regular dipolar chain uncoupled to cavity photons, since the light-matter detuning is too great to significantly influence the dipolar modes. For super-critical cavities ($L_x \ge L_{\mathrm{Tamm}}$), we observe the presence of Tamm-like edge states, which are characterized by a participation ratio which grows linearly with the cavity size for $L_x \gtrsim 8 a$. Explicitly, the dependency here is $\mathrm{PR}(m_{\mathrm{Tamm}}) \simeq 3.62 (L_x/a) + 5.39$, as shown by the dashed green line in panel (c) \cite{notelimit}. For smaller cavity sizes $L_{\mathrm{Tamm}} \le L_x \lesssim 8 a$, there is an interesting nonmonotonous behavior, and a global minimum of $\mathrm{PR}(m_{\mathrm{Tamm}}) \simeq 19.0$ occurs at $L_x = 5.64 a$ [cf. the results of Fig.~\ref{finiteeffect}~(a)]. Of course, these numbers depend on the chosen inter-dipole separation ratio $d/a$, dimensionless dipole strength $\omega_0 c/a$, and cavity aspect ratio $L_y/L_x$.

We have therefore demonstrated an unusual, non-topological [see the discussion after Eq.~\eqref{eq:Diraccy}] phase transition demarcating the absence and presence of Tamm-like edge states, which are induced by cavity interactions and boundary conditions in the chain beyond those used in the nearest-neighbor approximation. The observation of these proposed Tamm states requires careful sweeping in energy, due to their proximity to bulk states. Such careful measurements can be performed using the latest techniques in cathodoluminescence spectroscopy~\cite{Peng2019} and optical microscopy and spectroscopy~\cite{Mueller2020}. The detection of such states in the strong coupling regime provides perspectives for the fundamental understanding of the interplay between edge states and light-matter coupling, and for controlling the localization of polariton states in nanoscale waveguiding structures. Furthermore, while there is a great quest to find topological nanophotonic states~\cite{Lu2014, Khanikaev2017, Sun2017, Martinez2018, Ozawa2019, Rider2019, Ota2019}, our findings highlight that after the experimental observation of an edge state, one should also consider possible non-topological mechanisms of generation.


\section{Conclusion}
\label{conc}

We have presented a theory of polaritonic Tamm states, forged due to the mixing between the collective excitations in a dipolar chain and cavity photons. Importantly, the very existence of Tamm states requires the cavity cross-sectional area to be above a critical value. In this super-critical regime, the degree of localization of the Tamm states is highly dependent on the cavity size, with the participation ratio scaling linearly with the cavity cross-sectional area. We have also shown the crucial role played by dipole-dipole interactions beyond the celebrated nearest-neighbor approximation, without which the Tamm edge states do not form. The theory demonstrates the possibility of light trapping in non-topological one-dimensional structures, which may be important for waveguiding at the nanoscale. Our results also highlight how edge states may be generated via non-topological means, quite distinct from iconic topological models.

Our proposed model can be implemented in a host of systems based upon dipolar resonators, including dielectric and metallic nanoparticles~\cite{Chang2018}. Our theoretical proposal therefore offers the opportunity to finely control the propagation and localization of collective light-matter excitations at the subwavelength scale, and provides perspectives for more complicated and higher dimensional nanophotonic systems~\cite{Amico2019, Huang2020}.


\begin{acknowledgement}
  This work was supported by the the Arag\'{o}n government through the project Q-MAD. CAD acknowledges support from the Juan de la Cierva program (MINECO, Spain) and LMM was supported by the MINECO (Contract No. MAT2017-88358-C3-I-R).
\end{acknowledgement}




\begin{thebibliography}{100}



\bibitem{Tamm1932}
I.~E.~Tamm,
On the possible bound states of electrons on a crystal surface,
Phys. Z. Sowjetunion \textbf{1}, 733 (1932).

\bibitem{Fowler1933}
R.~H.~Fowler,
Notes on some electronic properties of conductors and insulators,
\href{https://doi.org/10.1098/rspa.1933.0103}
{Proc. Roy. Soc. A \textbf{141}, 56 (1933)}.

\bibitem{Shockley1939}
W.~Shockley,
On the surface states associated with a periodic potential,
\href{https://doi.org/10.1103/PhysRev.56.317}
{Phys. Rev. \textbf{56}, 317 (1939)}.

\bibitem{Forstmann1993}
F.~Forstmann,
The concepts of surface states,
\href{https://doi.org/10.1016/0079-6816(93)90055-Z}
{Prog. Surf. Sci. \textbf{42}, 21 (1993)}.

\bibitem{Davison1996}
S.~G.~Davison and M.~Steslicka,
\textit{Basic Theory of Surface States} (Oxford University Press, Oxford, 1996).

\bibitem{Ren2005}
S.~Y.~Ren,
\textit{Electronic States in Crystals of Finite Size: Quantum confinement of Bloch waves} (Springer, Berlin, 2005)

\bibitem{Ohno1990}
H.~Ohno, E.~E.~Mendez, J.~A.~Brum, J.~M.~Hong, F.~Agull\'{o}-Rueda, L.~L.~Chang, and L.~Esaki,
Observation of ``Tamm states'' in superlattices,
\href{https://doi.org/10.1103/PhysRevLett.64.2555}
{Phys. Rev. Lett. \textbf{64}, 2555 (1990)}.

\bibitem{Ohno1990b}
H.~Ohno, E.~E.~Mendez, A.~Alexandrou, and J.~M.~Hong,
Tamm states in superlattices,
\href{https://doi.org/10.1016/0039-6028(92)91112-O}
{Surf. Sci. \textbf{267}, 161 (1990)}.

\bibitem{Agranovich2000}
V.~M.~Agranovich, K.~Schmidt, and K.~Leo,
Surface states in molecular chains with strong mixing of Frenkel and charge-transfer excitons,
\href{https://doi.org/10.1016/S0009-2614(00)00678-3}
{Chem. Phys. Lett. \textbf{325}, 308 (2000)}.

\bibitem{Schmidt2002}
K.~Schmidt,
Quantum confinement in linear molecular chains with strong mixing of Frenkel and charge-transfer excitons,
\href{https://doi.org/10.1016/S0375-9601(01)00841-6}
{Phys. Lett. A \textbf{293}, 83 (2002)}.

\bibitem{Hoffmann2003}
M.~Hoffmann,
Mixing of Frenkel and charge-transfer excitons and their quantum confinement in thin films,
\href{https://doi.org/10.1016/S1079-4050(03)31005-1}
{Thin Films and Nanostructures \textbf{31}, 221 (2003)}.

\bibitem{Vinogradov2010}
For a review of Tamm states in photonic crystals, see
A.~P.~Vinogradov, A.~V.~Dorofeenko, A.~M.~Merzlikin and A.~A.~Lisyansky,
Surface states in photonic crystals,
\href{https://doi.org/10.3367/UFNe.0180.201003b.0249}
{Phys.-Usp. \textbf{53}, 243 (2010)}.

\bibitem{Kavokin2005}
A.~Kavokin, I.~Shelykh, and G.~Malpuech,
Optical Tamm states for the fabrication of polariton lasers,
\href{https://doi.org/10.1063/1.2136414}
{Appl. Phys. Lett. \textbf{87}, 261105 (2005)}.

\bibitem{Kaliteevski2007}
M.~Kaliteevski, I.~Iorsh, S.~Brand, R.~A.~Abram, J.~M.~Chamberlain, A.~V.~Kavokin, and I.~A.~Shelykh,
Tamm plasmon-polaritons: Possible electromagnetic states at the interface of a metal and a dielectric Bragg mirror,
\href{https://doi.org/10.1063/1.2136414}
{Phys. Rev. B \textbf{76}, 165415 (2007)}.

\bibitem{Savin2010}
A.~V.~Savin and Y.~S.~Kivshar,
Vibrational Tamm states at the edges of graphene nanoribbons,
\href{https://doi.org/10.1103/PhysRevB.81.165418}
{Phys. Rev. B \textbf{81}, 165418, (2010)}.


\bibitem{Goto2008}
T.~Goto, A.~V.~Dorofeenko, A.~M.~Merzlikin, A.~V.~Baryshev, A.~P.~Vinogradov, M.~Inoue, A.~A.~Lisyansky, and A.~B.~Granovsky,
Optical Tamm states in one-dimensional magnetophotonic structures,
\href{https://doi.org/10.1103/PhysRevLett.101.113902}
{Phys. Rev. Lett. \textbf{101}, 113902 (2008)}.


\bibitem{Nunez2016}
S.~N\'{u}\~{n}ez-S\'{a}nchez, M.~Lopez-Garcia, M.~M.~Murshidy, A.~G.~Abdel-Hady, M.~Serry, A.~M.~Adawi, J.~G.~Rarity, R.~Oulton, and W.~L.~Barnes,
Excitonic optical Tamm states: A step toward a full molecular-dielectric photonic integration,
\href{https://doi.org/10.1021/acsphotonics.6b00060}
{ACS Photonics \textbf{3} 743 (2016)}.

\bibitem{Fecteau2018}
A.~Juneau-Fecteau and L.~G.~Fr\'{e}chette,
Tamm plasmon-polaritons in a metal coated porous silicon photonic crystal,
\href{https://doi.org/10.1364/OME.8.002774}
{Opt. Mater. Express \textbf{8}, 2774 (2018)}.

\bibitem{Fecteau2019}
A.~Juneau-Fecteau, R.~Savin, A.~Boucherif, and L.~G.~Fr\'{e}chette,
Tamm phonon-polaritons: Localized states from phonon-light interactions,
\href{https://doi.org/10.1063/1.5089693}
{Appl. Phys. Lett. \textbf{114}, 141101 (2019)}.

\bibitem{Nakata2020}
Y.~Nakata, Y.~Ito, Y.~Nakamura, and R.~Shindou,
Topological boundary modes from translational deformations,
\href{https://doi.org/10.1103/PhysRevLett.124.073901}
{Phys. Rev. Lett. \textbf{124}, 073901 (2020)}.


\bibitem{Slobo2015}
A. P. Slobozhanyuk, A. N. Poddubny, A. E. Miroshnichenko, P. A. Belov, and Y. S. Kivshar, 
Subwavelength topological edge states in optically resonant dielectric structures, 
\href{http://dx.doi.org/10.1103/PhysRevLett.114.123901}
{Phys. Rev. Lett. \textbf{114}, 123901 (2015)}.

\bibitem{Ling2015}
C.~W.~Ling, M.~Xiao, C.~T.~Chan, S.~F.~Yu, and K.~H.~Fung,
Topological edge plasmon modes between diatomic chains of plasmonic nanoparticles,
\href{https://doi.org/10.1364/OE.23.002021}
{Opt. Express \textbf{23}, 2021 (2015)}.

\bibitem{Downing2017}
C.~A.~Downing and G.~Weick,
Topological collective plasmons in bipartite chains of metallic nanoparticles,
\href{https://doi.org/10.1103/PhysRevB.95.125426}
{Phys. Rev. B \textbf{95}, 125426 (2017)}.

\bibitem{Pocock2018}
S.~R.~Pocock, X. Xiao, P. A. Huidobro, and V. Giannini,
Topological plasmonic chain with retardation and radiative effects,
\href{https://doi.org/10.1021/acsphotonics.8b00117}
{ACS Photonics \textbf{5}, 2271 (2018)}.

\bibitem{Downing2018}
C.~A.~Downing and G.~Weick,
Topological plasmons in dimerized chains of nanoparticles: robustness against long-range quasistatic interactions and retardation effects,
\href{https://doi.org/10.1140/epjb/e2018-90199-0}
{Eur. Phys. J. B \textbf{91}, 253 (2018)}.

\bibitem{Gutierrez2018}
\'{A}. Guti\'{e}rrez-Rubio, L.~Chirolli, L.~Mart\'{i}n-Moreno, F.~J.~Garc\'{i}a-Vidal, and F.~Guinea,
Polariton anomalous Hall effect in transition-metal dichalcogenides,
\href{https://doi.org/10.1103/PhysRevLett.121.137402}
{Phys. Rev. Lett. \textbf{121}, 137402 (2018)}. 

\bibitem{Henriques2020}
J.~C.~G.~Henriques, T.~G.~Rappoport, Y.~V.~Bludov, M.~I.~Vasilevskiy, and N.~M.~R.~Peres,
Topological photonic Tamm states and the Su-Schrieffer-Heeger model,
\href{https://doi.org/10.1103/PhysRevA.101.043811}
{Phys. Rev. A \textbf{101}, 043811 (2020)}.    

\bibitem{Lu2014}
L.~Lu, J.~D.~Joannopoulos, and M.~Solja\v{c}i\'c, 
Topological photonics, 
\href{http://dx.doi.org/10.1038/NPHOTON.2014.248}
{Nat. Photon. \textbf{8}, 821 (2014)}.

\bibitem{Khanikaev2017}
A.~B.~Khanikaev and G.~Shvets, 
Two-dimensional topological photonics, 
\href{https://doi.org/10.1038/s41566-017-0048-5}
{Nat. Photon. \textbf{11}, 763 (2017)}.

\bibitem{Sun2017}
X.-C.~Sun, C.~H.~Xiao, P.~Liu, M.-H.~Lu, S.-N.~Zhu, and Y.-F.~Chen, 
Two-dimensional topological photonic systems, 
\href{https://doi.org/10.1016/j.pquantelec.2017.07.004}
{Prog. Quantum. Electron. \textbf{55}, 52 (2017)}.

\bibitem{Martinez2018}
V.~M.~Martinez~Alvarez, J.~E.~Barrios~Vargas, M.~Berdakin, and L.~E.~F.~Foa~Torres,
Topological states of non-Hermitian systems,
\href{https://doi.org/10.1140/epjst/e2018-800091-5}
{Eur. Phys. J. Spec. Top. \textbf{227}, 1295 (2018)}.

\bibitem{Ozawa2019}
T.~Ozawa, H.~M.~Price, A.~Amo, N.~Goldman, M.~Hafezi, L.~Lu, M.~C.~Rechtsman, D.~Schuster, J.~Simon, O.~Zilberberg, and I.~Carusotto,
Topological photonics,
\href{https://doi.org/10.1103/RevModPhys.91.015006}
{Rev. Mod. Phys. \textbf{91}, 015006 (2019)}.

\bibitem{Rider2019}
M.~S.~Rider, S.~J.~Palmer, S.~R.~Pocock, X.~Xiao, P.~A.~Huidobro, and V.~Giannini,
A perspective on topological nanophotonics,
\href{https://doi.org/10.1063/1.5086433}
{J. Appl. Phys. \textbf{125}, 120901 (2019)}.

\bibitem{Ota2019}
Y.~Ota, K.~Takaka, T.~Ozawa, A.~Amo, Z.~Jia, B.~Kante, M.~Notomi, Y.~Arakawa, and S.~Iwamoto,
Active topological photonics,
\href{https://doi.org/10.1515/nanoph-2019-0376}
{Nanophotonics \textbf{9}, 547 (2020)}.

\bibitem{Jean2017}
P.~St-Jean, V~ Goblot, E.~Galopin, A.~Lema\^{i}tre, T.~Ozawa, L.~Le~Gratiet, I.~Sagnes, J.~Bloch, and A.~Amo,
Lasing in topological edge states of a one-dimensional lattice,
\href{https://doi.org/10.1038/s41566-017-0006-2}
{Nat. Photonics \textbf{11}, 651 (2017)}.

\bibitem{Whittaker2019}
C.~E.~Whittaker, E.~Cancellieri, P.~M.~Walker, B.~Royall, L.~E.~Tapia~Rodriguez, E.~Clarke, D.~M.~Whittaker, H.~Schomerus, M.~S.~Skolnick, and D.~N.~Krizhanovskii,
Effect of photonic spin-orbit coupling on the topological edge modes of a Su-Schrieffer-Heeger chain,
\href{https://doi.org/10.1103/PhysRevB.99.081402}
{Phys. Rev. B \textbf{99}, 081402(R) (2019)}.



\bibitem{Brongersma2000}
M.~L.~Brongersma, J.~W.~Hartman, and H.~A.~Atwater,
Electromagnetic energy transfer and switching in nanoparticle chain arrays below the diffraction limit,
\href{https://doi.org/10.1103/PhysRevB.69.125418}
{Phys. Rev. B \textbf{62}, R16356(R) (2000)}.

\bibitem{Maier2003}
S.~A.~Maier, P.~G.~Kik, and H.~A.~Atwater,
Optical pulse propagation in metal nanoparticle chain waveguides,
\href{https://doi.org/10.1103/PhysRevB.67.205402}
{Phys. Rev. B \textbf{67}, 205402 (2003)}.

\bibitem{Park2004}
S.~Y.~Park and D.~Stroud,
Surface-plasmon dispersion relations in chains of metallic nanoparticles: An exact quasistatic calculation,
\href{https://doi.org/10.1103/PhysRevB.69.125418}
{Phys. Rev. B \textbf{69}, 125418 (2004)}.


\bibitem{Brandstetter2016}
A.~Brandstetter-Kunc, G.~Weick, C.~A.~Downing, D.~Weinmann, and R.~A.~Jalabert,
Nonradiative limitations to plasmon propagation in chains of metallic nanoparticles,
\href{https://doi.org/10.1103/PhysRevB.94.205432}
{Phys. Rev. B \textbf{94}, 205432 (2016)}.

\bibitem{Downing2018b}
C.~A.~Downing, E.~Mariani, and G.~Weick,
Retardation effects on the dispersion and propagation of plasmons in metallic nanoparticle chains,
\href{https://doi.org/10.1088/1361-648X/aa9d59}
{J. Phys.: Condens. Matter \textbf{30}, 025301 (2018)}.



\bibitem{Ameling2013}
R.~Ameling and H.~Giessen,
Microcavity plasmonics: strong coupling of photonic cavities and plasmons,
\href{https://doi.org/10.1002/lpor.201100041}
{Laser Photonics Rev. \textbf{7}, 141 (2013)}.

\bibitem{Ginzburg2016}
P.~Ginzburg,
Cavity quantum electrodynamics in application to plasmonics and metamaterials,
\href{https://doi.org/10.1016/j.revip.2016.07.001}
{Rev. Phys. \textbf{1}, 120 (2016)}.

\bibitem{Savelev2016}
R.~S.~Savelev, A.~V.~Yulin, A.~E.~Krasnok, Y.~S.~Kivshar,
Solitary waves in chains of high-index dielectric nanoparticles,
\href{https://doi.org/10.1021/acsphotonics.6b00384}
{ACS Photonics \textbf{3} 1869 (2016)}.

\bibitem{Bakker2017}
R.~M.~Bakker, Y.~Feng, Y.~R.~Paniagua-Dom\'{i}nguez, B.~Luk'yanchuk, A.~I.~Kuznetsov,
Resonant light guiding along a chain of silicon nanoparticles,
\href{https://doi.org/10.1021/acs.nanolett.7b00381}
{Nano Lett. \textbf{17} 3458 (2017)}.


\bibitem{Koshelev2020}
K.~Koshelev, S.~Kruk, E.~Melik-Gaykazyan, J.-H.~Choi, A.~Bogdanov, H.-G.~Park, and Y.~Kivshar,
Subwavelength dielectric resonators for nonlinear nanophotonics,
\href{https://doi.org/10.1126/science.aaz3985}
{Science \textbf{367} 288 (2020)}.

\bibitem{Barrow2014}
S.~J.~Barrow, D.~Rossouw, A.~M.~Funston, G.~A.~Botton, and P.~Mulvaney,
Mapping bright and dark modes in gold nanoparticle chains using electron energy loss spectroscopy,
\href{https://doi.org/10.1021/10.1021/nl5009053}
{Nano Lett. \textbf{14}, 3799 (2014)}.

\bibitem{Gur2018}
F.~N.~G\"{u}r, C.~P.~T.~McPolin, S.~Raza, M.~Mayer, D.~J.~Roth, A.~M.~Steiner, M.~L\"{o}ffler, A.~Fery, M.~L.~Brongersma, A.~V.~Zayats, T.~A.~F.~K\"{o}nig, and T.~L.~Schmidt,
DNA-assembled plasmonic waveguides for nanoscale light propagation to a fluorescent nanodiamond,
\href{https://doi.org/10.1021/acs.nanolett.8b03524}
{Nano Lett. \textbf{18}, 7323 (2018)}.

\bibitem{Rekola2018}
H.~T.~Rekola, T.~K.~Hakala, and P.~T\"{o}rm\"{a},
One-dimensional plasmonic nanoparticle chain lasers,
\href{https://doi.org/10.1021/acsphotonics.8b00001}
{ACS Photonics \textbf{5}, 1822 (2018)}.



\bibitem{Barth2010}
M.~Barth, S.~Schietinger, S.~Fischer, J.~Becker, N.~Nusse, T.~Aichele, B.~Lochel, C.~Sonnichsen, and O.~Benson,
Nanoassembled plasmonic-photonic hybrid cavity for tailored light--matter coupling,
\href{https://doi.org/10.1021/nl903555u}
{Nano Lett. \textbf{10},891 (2010)}.

\bibitem{Huang2011}
F.~M.~Huang, D.~Wilding, J.~D.~Speed, A.~E.~Russell, P.~N.~Bartlett, and J.~J.~Baumberg,
Dressing plasmons in particle-in-cavity architectures,
\href{https://doi.org/10.1021/nl104214c}
{Nano Lett. \textbf{11}, 1221 (2011)}.

\bibitem{Schmidt2012}
M.~A.~Schmidt, D.~Y.~Lei, L.~Wondraczek, V.~Nazabal, and S.~A.~Maier,
Hybrid nanoparticle--microcavity--based plasmonic nanosensors with improved detection resolution and extended remote-sensing ability,
\href{https://doi.org/10.1038/ncomms2109}
{Nat. Commun. \textbf{3}, 1108 (2012)}.

\bibitem{Schmidt2018}
Y.~Yin, J.~Wang, X.~Lu, Q.~Hao, E.~Saei, G.~Naz, C.~Cheng, L.~Ma, and O.~G.~Schmidt,
In situ generation of plasmonic nanoparticles for manipulating photon--plasmon coupling in microtube cavities,
\href{https://doi.org/10.1038/ncomms2109}
{ACS Nano \textbf{12}, 3726 (2018)}.


\bibitem{Pirmoradian2018}
F.~Pirmoradian, B.~Z.~Rameshti, M.~Miri, and S.~Saeidian,
Topological magnon modes in a chain of magnetic spheres,
\href{https://doi.org/10.1103/PhysRevB.98.224409}
{Phys. Rev. B \textbf{98}, 224409 (2018)}.

\bibitem{Zhang2014}
X.~Zhang, C.-L.~Zou, L.~Jiang, and H.~X.~Tang,
Strongly coupled magnons and cavity microwave photons,
\href{https://doi.org/10.1103/PhysRevLett.113.156401}
{Phys. Rev. B \textbf{113}, 156401 (2014)}.

\bibitem{Browaeys2016}
A.~Browaeys, D.~Barredo and T.~Lahaye,
Experimental investigations of dipole-dipole interactions between a few Rydberg atoms,
\href{https://doi.org/10.1088/0953-4075/49/15/152001}
{J. Phys. B: At. Mol. Opt. Phys. \textbf{49}, 152001 (2016)}.

\bibitem{Leseleuc2019}
S.~de~L\'{e}s\'{e}leuc, V.~Lienhard, P.~Scholl, D.~Barredo, S.~Weber, N.~Lang, H.~P.~B\"{u}chler, T.~Lahaye1, and A. Browaeys,
Observation of a symmetry-protected topological phase of interacting bosons with Rydberg atoms,
\href{https://doi.org/10.1126/science.aav9105}
{Science \textbf{365}, 775 (2019)}.

\bibitem{Weimer2012}
H.~Weimer, N.~Y.~Yao, C.~R.~Laumann, and M.~D.~Lukin,
Long-range quantum gates using dipolar crystals,
\href{https://doi.org/10.1103/PhysRevLett.108.100501}
{Phys. Rev. Lett. \textbf{108}, 100501 (2012)}.

\bibitem{Cooper2019}
N.~R.~Cooper, J.~Dalibard, and I.~B.~Spielman,
Topological bands for ultracold atoms,
\href{https://doi.org/10.1103/RevModPhys.91.015005}
{Rev. Mod. Phys. \textbf{91}, 015005 (2019)}.


\bibitem{Mann2018}
C.~R.~Mann, T.~J.~Sturges, G.~Weick, W.~L.~Barnes, and E.~Mariani,
Manipulating type-I and type-II Dirac polaritons in cavity-embedded honeycomb metasurfaces,
\href{https://doi.org/10.1038/s41467-018-03982-7}
{Nat. Commun. \textbf{9}, 2194 (2018)}.

\bibitem{Craig1984}
D.~P.~Craig and T.~Thirunamachandran,
\textit{Molecular Quantum Electrodynamics: An Introduction to Radiation-Molecule Interactions} (Academic Press, London, 1984).

\bibitem{Salam2010}
A.~Salam,
\textit{Molecular Quantum Electrodynamics: Long-Range Intermolecular Interactions} (Wiley, New Jersey, 2010).


\bibitem{Downing2019}
C.~A.~Downing, T.~J.~Sturges, G.~Weick, M.~Stobi\'{n}ska, and L.~Mart\'{i}n-Moreno,
Topological phases of polaritons in a cavity waveguide,
\href{https://doi.org/10.1103/PhysRevLett.123.217401}
{Phys. Rev. Lett. \textbf{123}, 217401 (2019)}.

\bibitem{SuppMatt}
See the Supplemental Material, which contains Refs.~\cite{Mariani2017}-\cite{Downing2019b} for further descriptions, details, and derivations.

\bibitem{Mariani2017}
C.~A.~Downing, E.~Mariani, and G.~Weick,
Radiative frequency shifts in nanoplasmonic dimers,
\href{https://doi.org/10.1103/PhysRevB.96.155421}
{Phys. Rev. B \textbf{96}, 155421 (2017)}.


\bibitem{Hopfield1958}
J.~J.~Hopfield,
Theory of the contribution of excitons to the complex dielectric constant of crystals,
\href{https://doi.org/10.1103/PhysRev.112.1555}
{Phys. Rev. \textbf{112}, 1555, (1958)}.


\bibitem{Downing2019b}
C.~A.~Downing, J.~C.~L\'{o}pez Carre\~{n}o, A.~I.~Fern\'{a}ndez-Dom\'{i}nguez, and E.~del Valle,
Asymmetric coupling between two quantum emitters,
\href{https://doi.org/10.1103/PhysRevA.102.013723}
{Phys. Rev. A \textbf{102}, 013723 (2020)}.

\bibitem{Milonni1994}
P.~W.~Milonni,
\textit{The Quantum Vacuum: An Introduction to Quantum Electrodynamics} (Academic Press, London, 1994).

\bibitem{Kakazu1994}
K.~Kakazu and Y.~S.~Kim,
Quantization of electromagnetic fields in cavities and spontaneous emission,
\href{https://doi.org/10.1103/PhysRevA.50.1830}
{Phys. Rev. A  \textbf{50}, 1830 (1994)}.



\bibitem{Berry1984}
M.~V.~Berry,
Quantal phase factors accompanying adiabatic changes,
\href{https://doi.org/10.1098/rspa.1984.0023}
{Proc. R. Soc. A \textbf{392}, 45 (1984)}.

\bibitem{Zak1989}
J.~Zak,
Berry's phase for energy bands in solids,
\href{http://dx.doi.org/10.1103/PhysRevLett.62.2747}
{Phys. Rev. Lett. \textbf{62}, 2747 (1989)}.

\bibitem{Asboth2016}
J.~K.~Asboth, L.~Oroszlany, and A.~Palyi,
\textit{A Short Course on Topological Insulators} (Springer, Heidelberg, 2016).

\bibitem{Bell1970}
R.~J.~Bell and P.~Dean,
Atomic vibrations in vitreous silica,
\href{https://doi.org/10.1039/DF9705000055}
{Discuss. Faraday Soc. \textbf{50}, 55 (1970)}.

\bibitem{Thouless1974}
D.~J.~Thouless,
Electrons in disordered systems and the theory of localization,
\href{https://doi.org/10.1016/0370-1573(74)90029-5}
{Phys. Rep. \textbf{13}, 93 (1974)}.

\bibitem{notelimit}
Of course, for very large cavities where $L_x \gg a$, any Tamm-like edge states are lost as they merge into the bulk part of the spectrum. This limit recovers the regular dipolar chain (uncoupled to light) which exhibits the complete absence of any highly localized states. However, this weak light-matter coupling limit goes beyond our strong coupling model, as encapsulated by the $2 \times 2$ Hamiltonian of Eq.~\eqref{eq:Ham_polly_other}. A proper treatment of the problem in the regime $L_x \gtrsim 15 a$ requires many photonic bands to be included in the theory, as discussed in Ref.~\cite{SuppMatt}.

\bibitem{Peng2019}
S.~Peng, N.~J.~Schilder, X.~Ni, J.~van~de~Groep, M.~L.~Brongersma, A.~Alu, A.~B.~Khanikaev, H.~A.~Atwater, and A.~Polman,
Probing the band structure of topological silicon photonic lattices in the visible spectrum,
\href{https://doi.org/10.1103/PhysRevLett.122.117401}
{Phys. Rev. Lett. \textbf{122}, 117401 (2019)}.

\bibitem{Mueller2020}
N.~S.~Mueller, Y.~Okamura, B.~G.~M.~Vieira, S.~Juergensen, H.~Lange, E.~B.~Barros, F.~Schulz, and S.~Reich,
Deep strong light-matter coupling in  plasmonic nanoparticle crystals,
\href{https://doi.org/10.1038/s41586-020-2508-1}
{Nature \textbf{583}, 780 (2020)}.

\bibitem{Chang2018}
D.~E.~Chang, J.~S.~Douglas, A.~Gonzalez-Tudela, C.-L.~Hung, and H.~J.~Kimble, 
Colloquium: Quantum matter built from nanoscopic lattices of atoms and photons,
\href{https://doi.org/10.1103/RevModPhys.90.031002}
{Rev. Mod. Phys. \textbf{90}, 031002 (2018)}.

 

\bibitem{Amico2019}
I.~D'Amico, D.~G.~Angelakis, F.~Bussieres, H.~Caglayan, C.~Couteau, T.~Durt, B.~Kolaric, P.~Maletinsky, W.~Pfeiffer, P.~Rabl, A.~Xuereb, and M.~Agio,
Nanoscale quantum optics,
\href{https://doi.org/10.1393/ncr/i2019-10158-0}
{Riv. Nuovo Cimento \textbf{42}, 153 (2019)}.

\bibitem{Huang2020}
L.~Huang, L.~Xu, M.~Woolley, and A.~E.~Miroshnichenko,
Trends in quantum nanophotonics,
\href{https://doi.org/10.1002/qute.201900126}
{Adv. Quantum Technol. \textbf{42}, 153 (2020)}.


\end{thebibliography}
\end{document}